\documentclass[12pt]{iopart}
\usepackage{epsf}   \usepackage{epsfig} \usepackage{color}

\newcommand{\ket}[1]{\mbox{$| #1 \rangle$}}

\begin{document}
\title{Fully-connected network of superconducting qubits in a cavity}

\author{Dimitris I. Tsomokos$^{1,2}$, Sahel Ashhab$^{2,3}$, Franco Nori$^{2,3}$}
\address{$^1$School of Physics, Astronomy \& Mathematics, Science
\& Technology Research Institute, University of Hertfordshire,
Hatfield AL10 9AB, UK}
\address{$^2$Advanced Science Institute, Institute of Physical and Chemical
Research (RIKEN), Wako-shi, Saitama 351-0198, Japan}
\address{$^3$Center for Theoretical Physics, Physics Department,
Applied Physics Program, Center for the Study of Complex Systems,
The University of Michigan, Ann Arbor, Michigan 48109-1040, USA}
\ead{d.tsomokos@gmail.com, ashhab@riken.jp, fnori@riken.jp}
\date{\today}

\begin{abstract}
A fully-connected qubit network is considered, where every qubit
interacts with every other one. When the interactions between the
qubits are homogeneous, the system is a special case of the finite
Lipkin-Meshkov-Glick model. We propose a natural implementation of
this model using superconducting qubits in state-of-the-art
circuit QED. The ground state, the low-lying energy spectrum and
the dynamical evolution are investigated. We find that, under
realistic conditions, highly entangled states of
Greenberger-Horne-Zeilinger and W types can be generated. We also
comment on the influence of disorder on the system and discuss the
possibility of simulating complex quantum systems, such as
Sherrington-Kirkpatrick spin glasses, with superconducting qubit
networks.
\end{abstract}

 \maketitle

\section*{Introduction}
The development of quantum information science has provided us
with a fresh perspective on condensed-matter physics. In the last
few years, new tools from quantum information theory have been
applied to several problems in many-body physics
\cite{Fazio_review}. At the center of this novel approach lies the
problem of entanglement, that is, how to quantify the genuinely
quantum correlations \cite{ent_review} in a many-body system and
what these correlations can tell us about the system itself. The
entanglement in a given system is also a resource for various
quantum information processing tasks. Spin chains and lattices
with short-range interactions have been studied extensively in
this context. Although interactions between neighbors are more
common, there are instances where long-range interactions give a
better description of physical systems, such as certain types of
spin glasses \cite{spin_glasses, Kirkpatrick_1989}. It would be
very interesting to be able to use quantum networks as simulators
of such complex quantum systems. Furthermore, connected networks
are attracting considerable interest in quantum information
science \cite{Acin_Cirac} and in many other fields
\cite{networks}.

Here we study a fully-connected network, where every qubit
interacts with every other one, irrespectively of the distances
between them. Crucially, the proposed model is readily
implementable with superconducting qubits \cite{Nori_review}. The
emerging field of circuit QED \cite{Nori_review, Chiorescu,
Wallraff,Johan} provides a natural system in which a large number
of qubits can be coupled together. In such systems,
superconducting qubits play the role of atoms, and a
harmonic-oscillator circuit element plays the role of a cavity
with which they interact. If a single `cavity' is simultaneously
coupled to a number of qubits, it will mediate coupling between
all the possible qubit pairs \cite{Ashhab}. If, in addition, the
cavity is far off resonance with the qubits, its degrees of
freedom can be integrated out of the problem and we obtain a
system in which all the qubits are pairwise interacting. Previous
studies have considered similar circuits for coupling arbitrarily
distant superconducting qubits \cite{previous_work}. However,
these studies relied on time-dependent pulses to selectively
couple one qubit pair at a time, whereas here we consider the
simultaneous coupling of all qubit pairs. An important incentive
for studying the fully-connected network is that all the different
elements for its construction are already in place in the
laboratory.

Next we introduce the model, study its low-lying energy spectrum
and its dynamical response, and discuss the influence of disorder.
Our analysis mainly concerns the entanglement properties of small
networks and the generation of highly-entangled states in
near-future experiments with existing technologies of
superconducting qubits in circuit QED. We also discuss the
possibility of simulating spin glasses with these systems.

\section*{Model and Hamiltonian}

To begin with, we consider $N$ charge (flux) qubits that are
coupled capacitively (inductively) and assume that each qubit is
operated at its degeneracy point~\cite{Nori_review}. Therefore the
Hamiltonian is
\begin{eqnarray} \label{H} {\cal H} = - \sum_{i=1}^{N}
\frac{\Delta_i}{2} Z_i - \sum_{(i,j)} J_{ij} X_i X_{j},
\end{eqnarray}
where the second sum runs over all possible qubit pairs. Here,
$\Delta_i > 0$ is the level splitting and $J_{ij}$ is the strength
of the coupling between qubits $i$ and $j$. $X_i$, $Y_i$ and $Z_i$
denote the Pauli matrices for qubit $i$. Multi-qubit entanglement
generation has been analyzed in trapped ions and atoms using
Hamiltonians of the form in Eq. (\ref{H}) \cite{ions,BE_sahel}. In
a circuit QED setup, which is the focus of this work, the
macroscopic qubits allow individual addressing and readout, in
addition to relatively straightforward scalability. Although
additional terms will appear in the Hamiltonian of this system,
these terms can be made negligibly small under realistic
conditions, as was shown in Ref.~\cite{Ashhab}.

If we let $\Delta_i = \Delta$ and $J_{ij} = J$ for all qubits, the
system is homogeneous and it corresponds to a special case of the
Lipkin-Meshkov-Glick model \cite{Fazio_review,LMG}. In this case,
the Hamiltonian can be expressed as
\begin{equation} \label{H-re-expressed}
{\cal H} = - \frac{\Delta}{2} Z_{\rm Total} - \frac{J}{2} X_{\rm
Total}^2 + \frac{NJ}{2},
\end{equation}
where $Z_{\rm Total}=\sum_{i=1}^{N} Z_i$ and $X_{\rm
Total}=\sum_{i=1}^{N} X_i$. From Eqs.~(\ref{H}) and
(\ref{H-re-expressed}) it is clear that the Hamiltonian commutes
with the square of the total pseudo-spin operator, $\sum_{i}(X_{i}
+ Y_{i} + Z_{i})^{2}$, and possesses spin-flip symmetry, i.e., it
also commutes with $\Pi_i Z_i$~\cite{LMG}.

\section*{Ground state properties}

Two parameter regimes can be identified, namely, $\Delta \gg N|J|$
and $\Delta \ll N|J|$. In the first case, the single-qubit term in
${\cal H}$ dominates and the preferred basis is the eigenbasis of
$Z$, $\{\ket{0}, \ket{1}\}$, with $Z \ket{0}=\ket{0}$ (the
eigenstates of $X$ are denoted by $\{\ket{+}, \ket{-}\}$). As $|J|
\rightarrow 0$ the ground state of the system becomes equal to the
fully-separable state
\begin{equation} \label{SEP}
\ket{\Psi_{\rm SEP}} = \ket{0}^{\otimes N}.
\end{equation}
We shall employ a pseudo-spin language for convenience; when
$J_{ij}<0$ we say that the interaction is `antiferromagnetic'
(AFM), and when $J_{ij}>0$ we say that the interaction is
`ferromagnetic' (FM).

Secondly, if $\Delta \ll N|J|$ then the interaction term in ${\cal
H}$ dominates and the preferred basis is the $\{\ket{+},
\ket{-}\}$ basis. In this case the sign of $J$ becomes important.
For large, positive $J$ (FM regime) the interaction term tends to
set all the qubits in the same state in the $\{\ket{+},\ket{-}\}$
basis. The `ideal' ground state of the system is approximately
\begin{equation} \label{GHZ}
\ket{\Psi_{\rm GHZ}} = \frac{1}{\sqrt{2}} \left(\ket{+}^{\otimes
N} + \ket{-}^{\otimes N} \right),
\end{equation}
which is known as the GHZ-state \cite{ent_review}. In practice,
however, this ideal state is typically fragile under small
external perturbations. By contrast, for large and negative $J$
(AFM regime) the interaction term favors a state in which pairs of
neighbouring qubits are antiparallel in the
$\left\{\ket{+},\ket{-}\right\}$ basis. Clearly, in a
fully-connected geometry this condition is impossible to satisfy
because every qubit neighbors every other qubit. This high degree
of frustration in the system leads to highly-entangled states, as
we show below. For the particular case of $N=3$ there is a
relatively simple ground state, namely,
\begin{equation} \label{ENT}
\ket{\Psi_{\rm ENT}} = \frac{1}{\sqrt{3}} \left( \ket{+-0} +
\ket{-0+} + \ket{0+-} \right).
\end{equation}

\begin{figure}
  \centering
  \resizebox{0.49\linewidth}{!}{\includegraphics{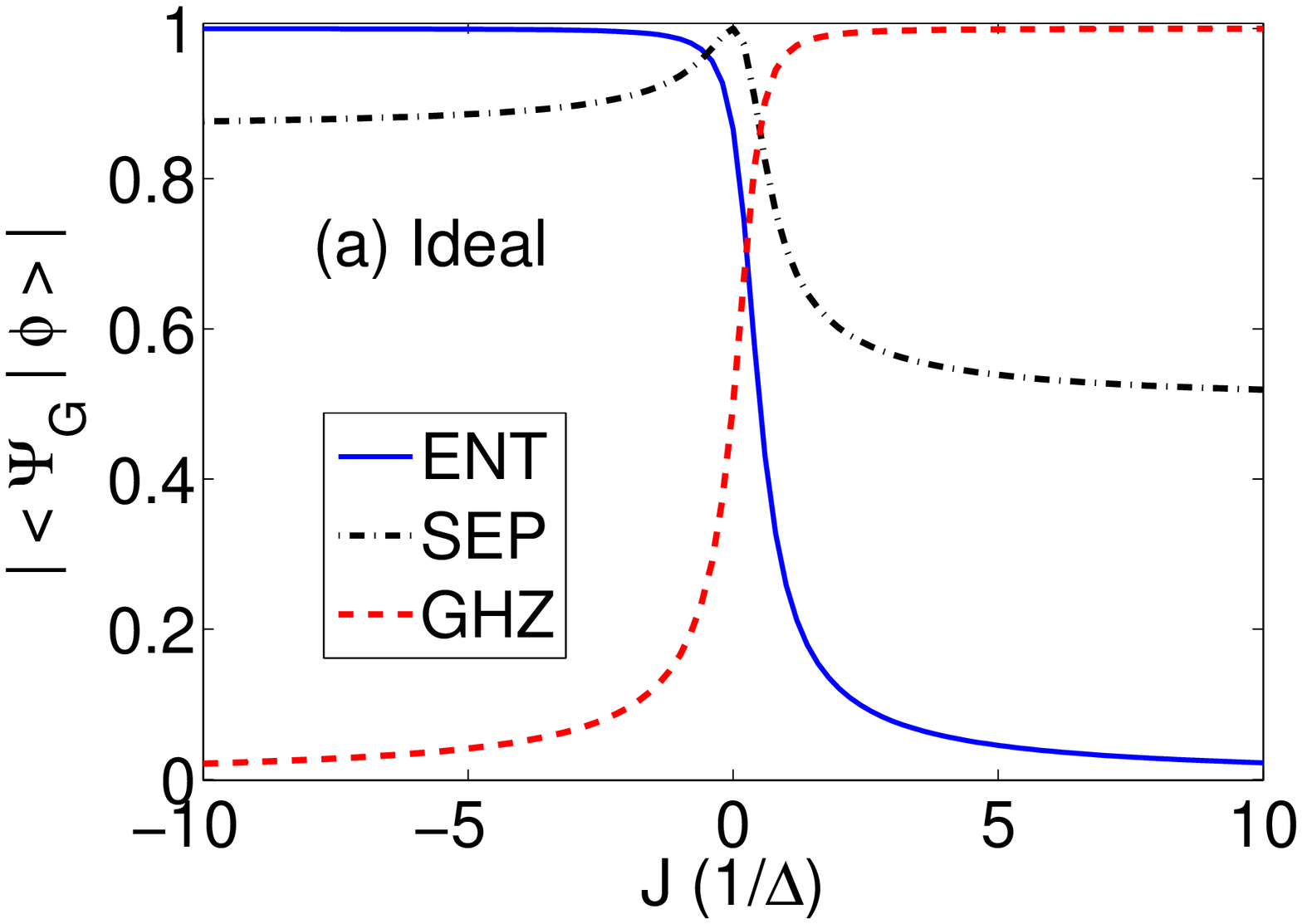}}
  \resizebox{0.49\linewidth}{!}{\includegraphics{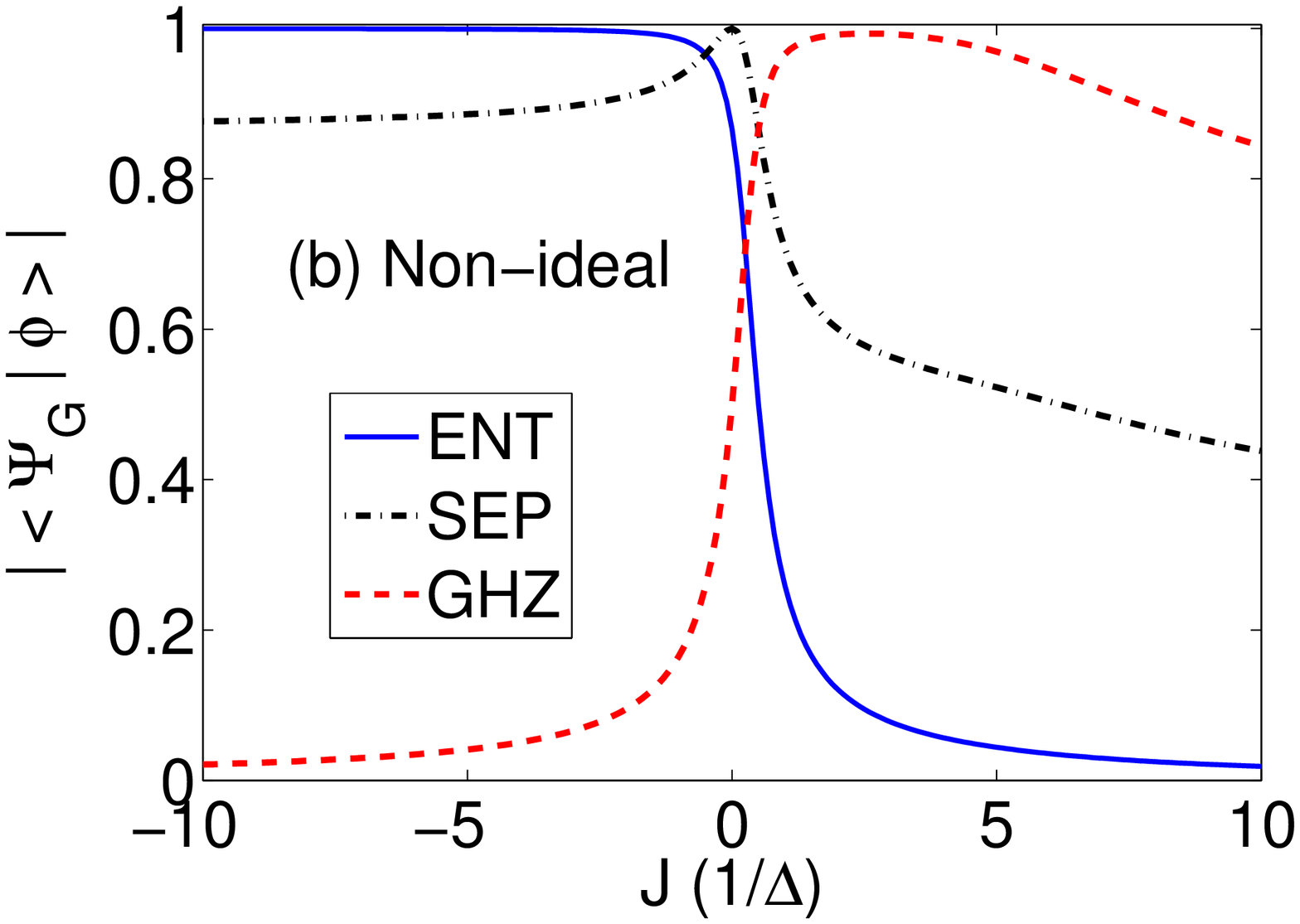}}
  \caption{Fidelity $|\langle \Psi_{\rm G} | \phi \rangle|$
  between the ground state, $\ket{\Psi_{\rm G}}$,
  and the states $\ket{\Psi_{\rm SEP}}$, $\ket{\Psi_{\rm GHZ}}$,
  $\ket{\Psi_{\rm ENT}}$ against $J$ for a network of $N=3$ qubits.
  We show (a) the ideal case and (b) the case where a small perturbation $g X_i$
  of strength $g = \Delta/100$ is applied to qubit $i$.}
  \label{fig__fidelities}
\end{figure}

We illustrate the above statements by means of an exact numerical
diagonalization, ${\cal H}\ket{\Psi_n}=E_n\ket{\Psi_n}$ (for
$n=1,2,\ldots$). In all numerical simulations, we let $\Delta = 1$
and hence express the results in units of $\Delta$. In
Fig.~\ref{fig__fidelities}(a) we show the fidelity $|\langle
\Psi_{\rm G} | \phi \rangle|$ between the ground state of the
system, $\ket{\Psi_{\rm G}}$, and the three states $\ket{\phi}$ of
Eqs. (\ref{SEP}) to (\ref{ENT}) for $N=3$, in the ideal case. In
Fig.~\ref{fig__fidelities}(b) we calculate the same fidelities but
now we apply a small perturbation $gX_i$, of strength $g \ll
\Delta$, to qubit $i$ (it does not matter which one). The
symmetry-breaking term only affects the FM regime, where the
$\ket{\Psi_{\rm GHZ}}$ becomes increasingly fragile as $J$ is
increased beyond a certain optimal value. This result holds for
all small networks with $N \sim 10$: it is always possible to find
an optimal value of the coupling strength such that the ground
state is very close to a GHZ-state, even in the presence of a
small external perturbation. For larger networks, or stronger
perturbations, the behaviour is more abrupt and we do not obtain
exact or almost exact GHZ-states, in practice.

\begin{figure}
  \centering
  \resizebox{0.49\linewidth}{!}{\includegraphics{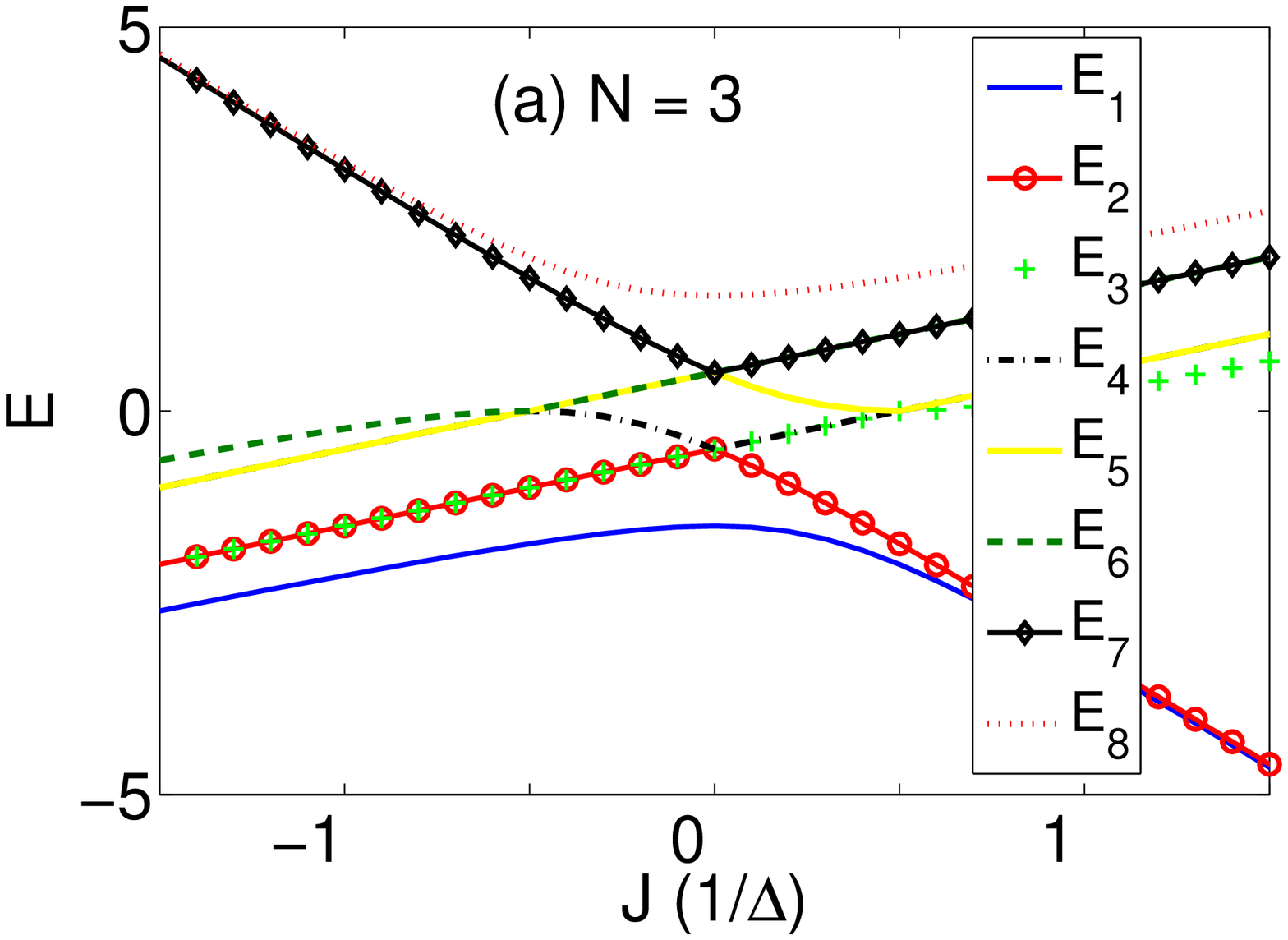}}
  \resizebox{0.49\linewidth}{!}{\includegraphics{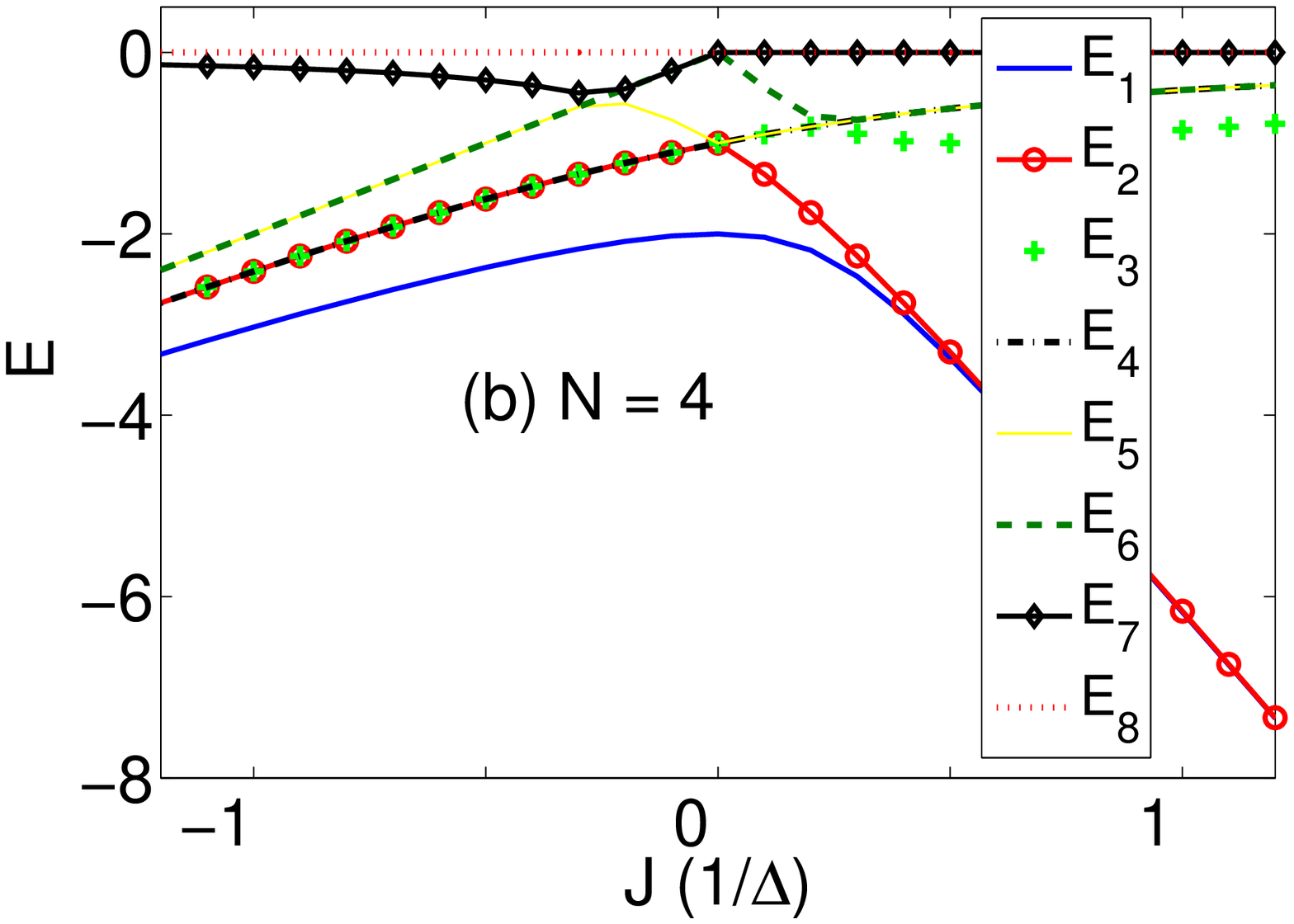}}
  \caption{(a) Full energy spectrum for a system of size $N=3$
  against $J$. (b) Lower half of the energy spectrum, versus $J$, for a system
  of size $N=4$.}
  \label{fig__spectra}
\end{figure}

In Fig.~\ref{fig__spectra}(a) the full energy spectrum for the
$N=3$ case is presented. In Fig.~\ref{fig__spectra}(b) the lower
half of the spectrum for $N=4$ is presented. One can see that the
ground state is unique in the AFM regime and two-fold degenerate
in the FM regime. Also, there is a finite energy gap between the
ground states and the first excited states, which remains constant
with increasing $|J|$ in the AFM regime and increases with $|J|$
in the FM regime.

We now turn to the entanglement properties of the different
possible ground states. We consider both the entanglement between
pairs of qubits and also between blocks of qubits. Among the
various measures of two-qubit entanglement \cite{ent_review}, we
calculate the logarithmic negativity \cite{log_negativity}. It is
defined as
\begin{equation} \label{log_neg}
E_{\rm N}(\rho_{ij}) \equiv \log_2 ||\rho_{ij}^{T_i}||,
\end{equation}
where $||.||$ denotes the trace norm of a matrix and
$\rho_{ij}^{T_i}$ is the partial transpose of the reduced density
matrix $\rho_{ij}$ of two qubits $i$ and $j$. We also use the von
Neumann entropy of a state $\rho_k$, for a block of $k<N$ qubits,
\begin{equation} \label{vonN_entropy}
S(\rho_k) \equiv - {\rm tr}(\rho_k \log_2 \rho_k).
\end{equation}
$S(\rho_k)$ quantifies how mixed is the reduced density matrix
$\rho_k$ and, if the system as a whole is in a pure state, it also
quantifies the amount of entanglement between the qubits in the
set $k$ and those in the rest of the system.

\begin{figure}
\centering
  \resizebox{0.49\linewidth}{!}{\includegraphics{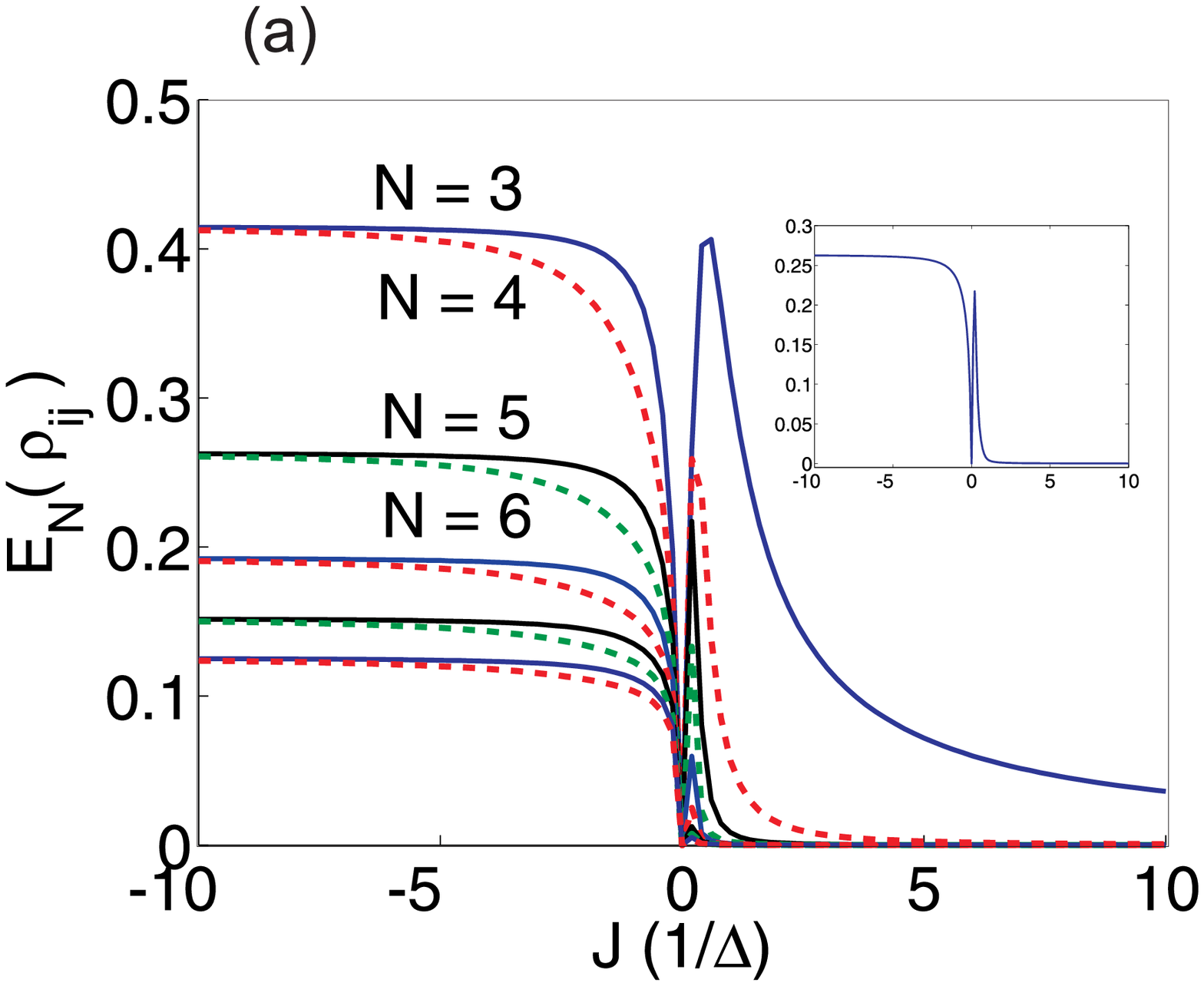}}
  \resizebox{0.49\linewidth}{!}{\includegraphics{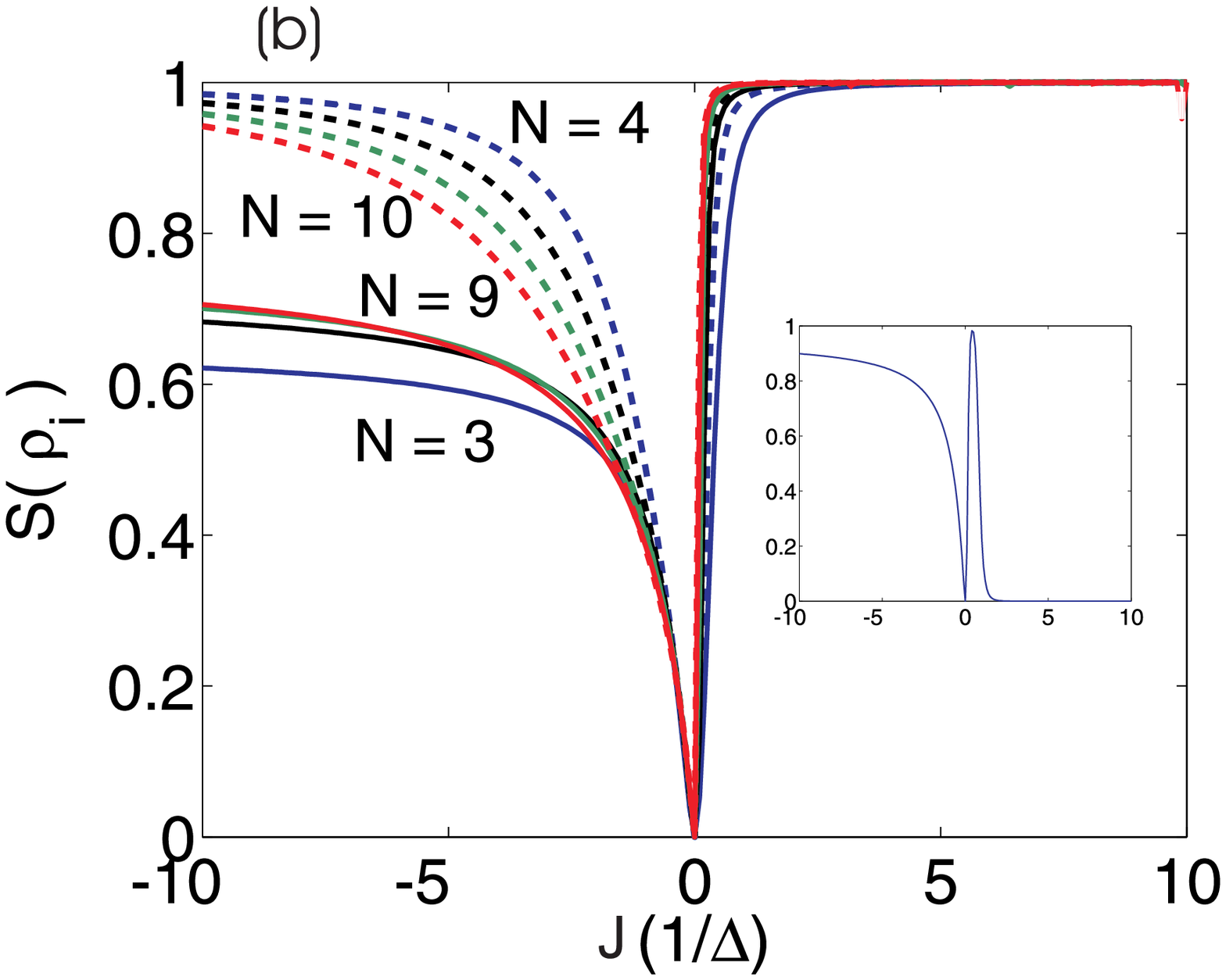}}
  \caption{(a) $E_{\rm N}(\rho_{ij})$ against $J$
  for networks of size $N=3$ to $N=12$ (from top to bottom).
  (b) $S(\rho_i)$ against $J$ for $N=3$ to $N=10$. Solid lines
  correspond to $N=3,5,7,9$ (from bottom to top) and broken
  lines to $N=4,6,8,10$ (from top to bottom), as indicated.
  In both (a) and (b), the insets correspond to the $N=5$ case
  with a small perturbation $g X_i$ of strength $g = \Delta/100$.
  }
  \label{fig__GS_ent}
\end{figure}

The pairwise entanglement, $E_{\rm N}(\rho_{ij})$, is shown in
Fig.~\ref{fig__GS_ent}(a) against $J$ for systems of size $N=3,4,
\ldots, 12$. We observe that $E_{\rm N}(\rho_{ij})$ decreases with
the size of the system $N$. This can be explained by the fact that
as $N$ increases so do the number of interactions for each
individual qubit. This higher degree of connectivity places
stronger monogamy constraints that generally reduce the two-party
entanglements \cite{CKW_2000}. To study the many-body correlations
of the ground state, we show in Fig.~\ref{fig__GS_ent}(b) the
entanglement of a single qubit $i$ with the rest of the system,
using the von Neumann entropy $S(\rho_i)$. In the FM regime the
ideal ground state is a GHZ state, in which every qubit is
maximally entangled with the rest of the network. In practice,
however, under the influence of an external perturbation, the
system chooses one of the two degenerate eigenstates and becomes
fully separable in the deep FM regime. This is seen in the inset
of Fig.~\ref{fig__GS_ent}(b), which shows the ground state of a
small network perturbed by $g X_i$ with strength $g = \Delta/100$.
In the AFM regime, $S(\rho_i)$ increases with the size of the
network for odd $N$ and it decreases for even $N$. This is due to
the different symmetries in the two cases, which appear for small
networks but disappear in the thermodynamic limit (see, e.g.,
\cite{LMG}). In summary, in the AFM regime the system achieves
multi-qubit entanglement for any $|J|>0$; in the FM regime the
system achieves a high degree of multi-qubit entanglement in
practice, approaching a GHZ-state, for an optimal value of the
interaction strength, while deep in the FM regime it becomes fully
separable.

\section*{Dynamical evolution}

Next, we study the system's evolution, which is determined by the
state
\begin{equation}
\ket{\Psi(t)}=\exp(-i{\cal H}t)\ket{\Psi(0)}.
\end{equation}
We consider two simple initial states: the state $\ket{\Psi_{\rm
SEP}}$ of Eq.~(\ref{SEP}), and the state $\ket{\Psi^{(1)}_{\rm
SEP}} = \ket{1}\otimes \ket{0}^{\otimes N-1}$, which is the same
as $\ket{\Psi_{\rm SEP}}$ except that the state of one qubit is
flipped. Both of these states are separable and easy to prepare
experimentally. For definiteness we assume that $J>0$, and we only
consider the case of weak coupling, which is most relevant to
near-future experiments.

In the weak-coupling limit, $J \ll \Delta / N$, one can classify
the low-lying states according to the number of elementary
excitations they contain. The ground state $\ket{\Psi_{\rm SEP}}$
contains no excitations and does not evolve in time. At an energy
$\sim \Delta$ above the ground state, there are $N$ energy
eigenstates that can be identified as one-excitation states. These
states have the `spin-wave' form $\ket{\Phi_{1,k}} =
\frac{1}{\sqrt{N}} \sum_{j=1}^{N} e^{2\pi ijk/N} \ket{0_1 \cdots
0_{j-1} 1_j 0_{j+1} \cdots 0_N}$, where $k=0,1,\cdots ,N-1$. Their
energies (relative to the ground state) are given by
$\Delta-(N-1)J$ for $k=0$ and $\Delta+J$ otherwise. As these $N$
energy eigenstates are separated from all other states by an
energy at least $\sim \Delta$, their dynamics can be analyzed in
the restricted Hilbert space containing only these $N$ states.
Using the above spectrum, we find that the initial state
$\ket{\Psi(t=0)}=\ket{\Psi^{(1)}_{\rm SEP}}$ evolves, up to an
overall phase factor, into
\begin{eqnarray} \label{Psi_t}
\ket{\Psi(t)} = \ket{\Psi^{(1)}_{\rm SEP}} + \frac{\exp(i N J
t)-1}{\sqrt{N}} \ket{W_N},
\end{eqnarray}
where the `W-state' \cite{ent_review} is $\ket{W_N} =
\frac{1}{\sqrt{N}} \sum_{j=1}^{N} \ket{0_1 \cdots 0_{j-1} 1_j
0_{j+1} \cdots 0_N}.$ From this result we observe that the evolved
state is a time-dependent superposition of $\ket{\Psi^{(1)}_{\rm
SEP}}$ and $\ket{W_N}$. For large networks ($N \rightarrow
\infty$) the second term can be neglected, and the system remains
close to its initial state (the initial excitation is localized).
For $N=3$ and $N=4$, on the other hand, Eq.~(\ref{Psi_t}) reduces
to variants of the W-state (when $|e^{iNJt}-1|=\sqrt{N}$),
\begin{eqnarray} \label{W}
\ket{{\cal W}_{3}} & = & \frac{e^{\pm i\pi/6}}{\sqrt{3}} \ket{1 0
0} + \frac{e^{\pm 5i\pi/6}}{\sqrt{3}} \left( \ket{0 1 0} + \ket{0
0 1} \right)
\nonumber \\
\ket{{\cal W}_{4}} & = & \frac{1}{2} ( - \ket{1 0 0 0} + \ket{0 1
0 0} + \ket{0 0 1 0}  + \ket{0 0 0 1}).
\end{eqnarray}

\begin{figure}
  \centering
  \resizebox{0.49\linewidth}{!}{\includegraphics{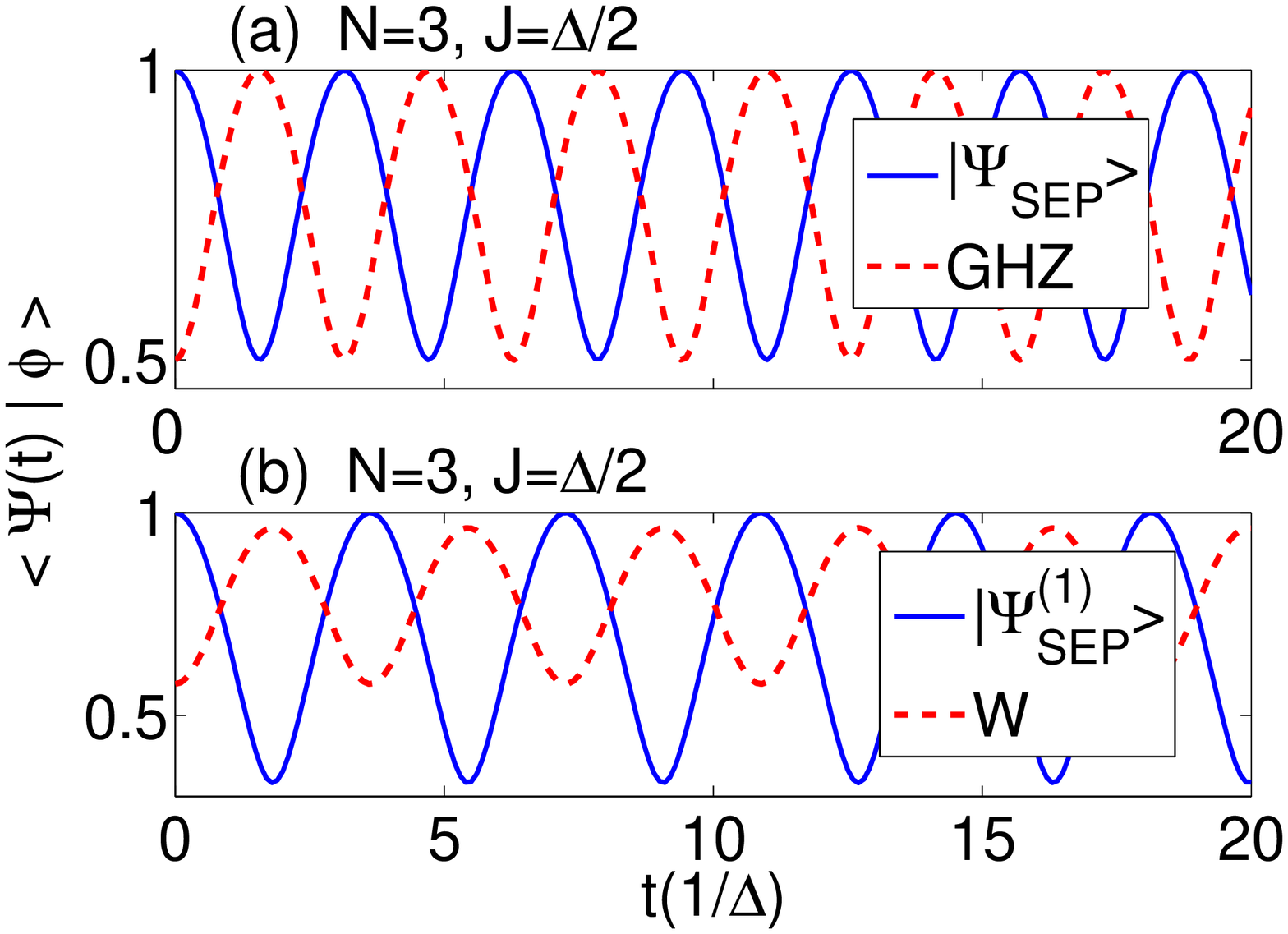}}
  \resizebox{0.49\linewidth}{!}{\includegraphics{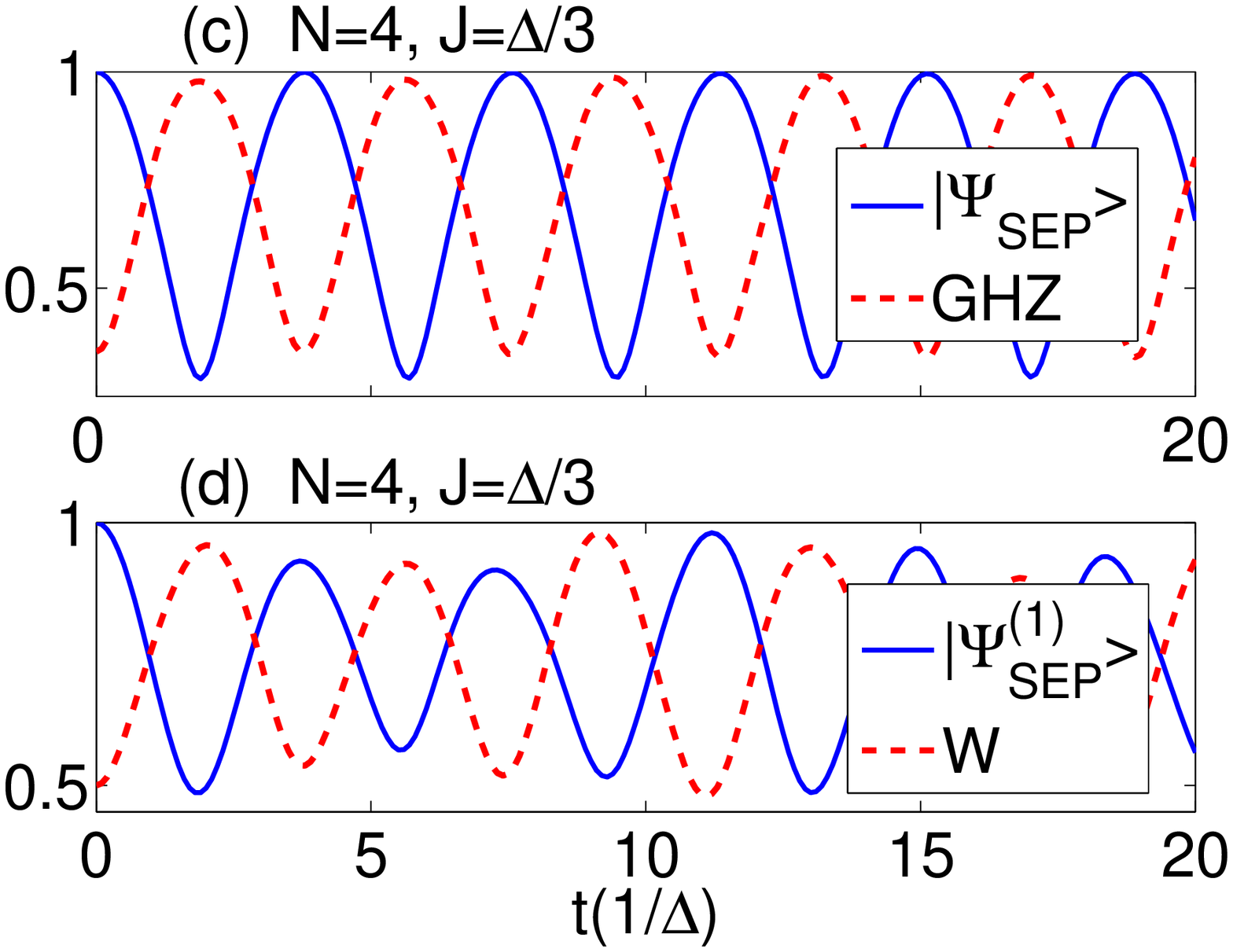}}
  \caption{Fidelity $|\langle \Psi(t)|\phi \rangle |$
  between the evolved state and the different initial
  states, the GHZ-state of Eq.~(\ref{GHZ}) and the W-state of Eq.~(\ref{W}),
  as indicated. (a) and (b) correspond to $N=3$, $J = \Delta/2$,
  while (c) and (d) correspond to $N=4$, $J = \Delta/3$.}
  \label{fig__Fidelity_Evolution}
\end{figure}

For larger, but still weak, coupling we study the evolution using
the fidelity between $\ket{\Psi(t)}$ and the GHZ and W states. If
the initial state is $\ket{\Psi_{\rm SEP}}$, then the system
evolves into a state that is close to a GHZ-state; if the initial
state is $\ket{\Psi^{(1)}_{\rm SEP}}$, then the system evolves
into a state that is close to a W-state, for small networks. For
$N=3$ and $N=4$, $\ket{\Psi(t)}$ can be made arbitrarily close to
$\ket{\Psi_{\rm GHZ}}$ or $\ket{{\cal W}}$ for properly chosen
values of $J$ (see Fig.~\ref{fig__Fidelity_Evolution}). It is
possible to achieve multi-qubit entanglement of these two types
with networks of size $N>4$, but the window for the required
coupling strengths is much sharper and the fidelity maxima
decrease below $1$. For instance, for $N=6$ and $0 < J < \Delta/2$
the maximum fidelity of the evolved state with the GHZ-state is
$0.96$, which is still a relatively high fidelity \cite{footnote}.

Thus, the fully connected network is well suited for the fast
(one-step) preparation of GHZ- and W- entangled states. This
result should be contrasted with recent work on the generation of
entangled states in superconducting qubit circuits using generally
long sequences of basic operations \cite{ENT_sc_qubits}. For
networks with $N>4$ and the simple initial states used above, we
obtain entangled states that are different from the GHZ- and
W-states, but still highly entangled and therefore of potential
value for future quantum information technologies (e.g., we only
mention applications related to quantum secret sharing and quantum
average estimation \cite{applications}).

Finally, we note that the dynamics of entanglement has simple
periodicity only in the case of $N=3$, which corresponds to a
closed chain. In general, for $N>3$, the behavior of the
logarithmic negativity $E_N(\rho_{ij})$ is complicated and there
are time intervals for which it drops to zero. As the system size
increases, so does the durations of these time intervals. During
most of the intervals in which the pairwise entanglements vanish,
the block entanglements increase to their local maxima. Therefore
the system achieves a form of multi-qubit entanglement. On the
other hand, for small and negative $J$ we find that the evolution
of entanglement shows more periodic features. The block entropies
$S(\rho_k)$ (for different block sizes $k$) have their local
maxima at the same times as the $E_N(\rho_{ij})$. Hence in the AFM
regime the system evolves into an entangled state that is
different from the GHZ- and W-states.

\section*{Static disorder}

In the presence of disorder, a system with the Hamiltonian ${\cal
H}(\Delta_i, J_{ij})$ of Eq.~(\ref{H}) is inhomogeneous. As a
result, different partitions of the network that correspond to the
same number of qubits are no longer equivalent. We study the
influence of uniform, static disorder on the system by assuming
that the $J_{ij}$ and the $\Delta_i$ are chosen randomly from the
intervals $[1-\delta_{1},1+\delta_{1}]J$ and $[1-\delta_{2},
1+\delta_{2}]\Delta$, respectively, where $\delta_{1,2} \in [0,1]$
quantify the amount of disorder in each parameter. We neglect the
type of disorder that can move a qubit away from its degeneracy
point, but this is a valid approximation for realistic
implementations. The properties of the low-lying energy sector and
the dynamics are studied as an average over many realizations of
${\cal H}(\Delta_i, J_{ij})$. We only consider the weak-coupling
regime, and the case of small networks.

We shall not show any details on these Monte-Carlo simulations, as
they do not add any insight beyond the main conclusions. From the
numerical calculations, we observe that disorder in the $J_{ij}$
is more important than disorder in the $\Delta_i$. More crucially,
we find that the results reported here, including those on the
ground state entanglement and the dynamics of entanglement, remain
largely unaffected for disorder of amount $\delta_{1,2} \le 10
\%$. This result is in agreement with previous studies on disorder
\cite{tsomokos}, and it is an experimentally achievable upper
bound.

\section*{Spin glasses}

The simplest examples of glassy systems are spin glasses and they
offer the possibility of studying the behavior of complex systems
away from equilibrium. Spin glasses arise when the interactions
between spins are ferromagnetic for some bonds and
antiferromagnetic for others, in which case the spin orientation
cannot be uniform in space even at low temperatures
\cite{spin_glasses}. In this case the spin orientation can become
random and frozen in time. A particularly illuminating and
extensively studied model of spin glasses is the
Sherrington-Kirkpatrick (SK) model. The SK model in a transverse
field \cite{Kirkpatrick_1989} is given by the Hamiltonian ${\cal
H}$ of Eq. (\ref{H}) for interactions that are disordered in both
magnitude and sign. In fact, if the distribution of $J_{ij}$ is
Gaussian, then for $J \sim \Delta $ the system is a spin glass for
temperatures lower than $k_{B} T \sim \Delta / 4$ (see, e.g.,
\cite{Kirkpatrick_1989}).

There are various important open problems in this field, both
theoretical and experimental. For instance, in relation to the
proposal of the present work, one such problem is the behavior of
spin glasses in the low temperature phase region, where quantum
phenomena dominate (see, e.g., Parisi in \cite{spin_glasses}).
Some of the advantages of using qubit networks as quantum
simulators include the fact that the states of all the qubits can
be prepared controllably and that the dynamics of all the qubits
can be monitored as a function of time, yielding the `microscopic'
dynamics of individual spins and not just averaged spin
quantities. Therefore the implementation of fully-connected
networks with superconducting qubits, which can be addressed and
measured individually, can offer valuable additional tools for the
study of complex quantum systems, such as spin glasses.

\section*{Outlook and summary}

In this work we have focused on $XX$-type interqubit couplings
[Eq.~(\ref{H})] since this is relevant to present-day experiments
in circuit QED. A modified version of the flux qubit was proposed
recently, which implements $ZZ$-type couplings \cite{Kerman}. This
form of the coupling could give rise to other types of many-body
entangled quantum states \cite{graphs}. We have also focused on
the case where the cavity is far off resonance with the qubits.
Bringing the cavity into resonance with the qubits would lead to a
star geometry with the cavity at the center of the star. In this
case, the Hamiltonian of Eq. (\ref{H}) is no longer valid;
instead, each qubit interacts only with the cavity and not other
qubits. The addition of a nonlinearity to the cavity can be used
to make the center of the star an effective two-level system
\cite{Bose}.

In conclusion, we have proposed an implementation of a qubit
network where all qubits are coupled in pairs, independently of
the relative distances between them, as in the finite LMG model of
spin systems. Such a network supports a highly-entangled ground
state in the case of antiferromagnetic interactions, which is
robust against small external perturbations. Under suitable
conditions, separable initial states evolve into exact GHZ- and
W-type states in the case of small networks, or other
highly-entangled states for larger networks. The qualitative
behavior of the system is unaffected by the presence of static
disorder, as long as the amount of disorder is under about $10\%$.
Thus the system is well-suited for the generation of many-body
entangled states with macroscopic superconducting qubits. The
presence of entanglement in such systems can be witnessed
experimentally via combinations of two-qubit correlation
measurements, as for example described in Ref. \cite{tsomokos}.
Another promising prospect is the simulation of spin glasses. The
fully-connected network could be realized experimentally in the
near future with superconducting qubits in circuit QED.

\ack
We would like to thank A. Galiautdinov for useful discussions.
This work was supported in part by the NSA, LPS, ARO, NSF (grant
No.~EIA-0130383) and the JSPS-CTC program. DIT acknowledges the
support of the EPSRC (EP/D065305/1).

\section*{References}

\end{document}